\documentclass[twocolumn,prc,showpacs,preprintnumbers,amsmath,amssymb]
{revtex4}
\usepackage{graphicx}
\usepackage{dcolumn}
\usepackage{bm}

\newcommand{\beq}{\begin{eqnarray}}
\newcommand{\eeq}{\end{eqnarray}}

\newcommand{\btem}{\bibitem}

\begin{document}


\title{ Simultaneous Softening of $\sigma$ and $\rho$ Mesons
        associated with Chiral Restoration}

\author{K. Yokokawa$^{(1)}$, T. Hatsuda$^{(1)}$, A. Hayashigaki$^{(1)}$,
T. Kunihiro$^{(2)}$}
\address{$^{(1)}$ Department of Physics, University of Tokyo, Tokyo
113-0077 Japan }
\address{$^{(2)}$  Yukawa Institute for Theoretical Physics, Kyoto
University,
Kyoto 606-8502, Japan }

\date{\today}

\begin{abstract}
      Complex poles of the unitarized
      $\pi$-$\pi$ scattering amplitude in nuclear matter
      are studied. Partial restoration of chiral symmetry
      is modeled by the decrease of  in-medium
      pion decay constant  $f^*_{\pi}$.
      For large chiral restoration ($f^*_{\pi}/f_{\pi} \ll 1$),
       2nd sheet poles in the scalar ($\sigma$) and
      the vector ($\rho$) mesons
      are both dictated by the Lambert $W$ function and show
      universal softening as $f^*_{\pi}$ decreases.
      In-medium $\pi$-$\pi$ cross section receives
      substantial contribution from the soft mode and
      exhibits a large enhancement in low-energy region.
      Fate of this universality for small chiral restoration
      ($f^*_{\pi}/f_{\pi} \sim 1$)   is also  discussed.
\end{abstract}


\maketitle


          An analogy between  in-medium hadrons in QCD and
          collective modes in condensed matters suggests that
          hadronic spectral functions and
          complex hadronic-poles have important information
          on chiral structure of hot/dense matter \cite{HKBR}.
         In particular,   the light
           vector mesons ($\rho, \omega, \phi$) \cite{PI-BR-HL},
           and  the light scalar meson ($\sigma$) \cite{HK85}
           have been proposed to be possible probes of
        partial restoration of chiral symmetry.
      Properties of in-medium vector mesons may be
          extracted  through   dilepton emission from
           hot and/or dense matter.
        In fact, relativistic heavy ion experiments
           in SPS at CERN  indicate some
           spectral broadening and/or shift of the $\rho$
        \cite{CERES,RAWA}.
           Dileptons observed  in  proton-nucleus reactions
          at KEK also suggest a low-mass spectral enhancement
         in the vector channel  \cite{E325}.

         The $\sigma$ meson, on the other hand,  has been thought
         to be a direct but difficult probe of chiral restoration
         because of its obscure nature in the vacuum.
         Nevertheless,
         there arise growing interests  in recent years to
         the $\sigma$ in nuclear matter inspired by
        new experimental data of the
         $\pi$-$\pi$ invariant mass
         distribution in $(\pi, \pi \pi)$   \cite{CHAOS,CB,grion}
        and  $(\gamma, \pi \pi)$ \cite{Mainz} reactions
         with nuclear targets \cite{TK95,Orsay}.
         They suggest some spectral
         change of the two-pion final states  in the
         $I$=$J$=0  channel ($I$ and $J$ being the total isospin and
        total angular momentum, respectively).
       Theoretically,   the $\sigma$ pole located deep
     in the complex  energy
         plane may move toward the real-axis \cite{HK85} and
         behave as an important precursory mode if
         the partial chiral restoration occurs \cite{HKS,JHK,Orsay}.

        The main aim  of this paper is to investigate
        a common
        mechanism dictating the behavior of both $\sigma$ and
        $\rho$ when the partial chiral restoration takes place
        in nuclear matter. For this purpose,
         we study  complex poles of the
          $\pi$-$\pi$ scattering amplitude in nuclear matter
        calculated in unitarized chiral models.
        To make reliable unitarization and  analytic continuation
        of the scattering amplitude
        to complex energies, we adopt the $N/D$ method \cite{ND,IH,OO}
         which   respects
        the analyticity  and  approximately satisfies the crossing
   symmetry.
      As for the chiral models to be unitarized, we adopt
        four complementary models   shown below. They  are
             useful to check  the model dependence of the final results.

\noindent
{\it Model A}: The ``$\rho$ model"  \cite{BZ,BAS} where
       $\pi$ and bare-$\rho$ are the basic degrees of freedom.
       The physical $\sigma$ is generated dynamically in this model.

\noindent
{\it Model B}: The  ``$\sigma$ model"   \cite{BL,BAS}
       where  $\pi$ and  bare-$\sigma$ are the basic degrees of freedom.
        The physical $\rho$ is  generated dynamically in this model.

\noindent
{\it Model C}: The ``degenerate $\rho$-$\sigma$ model"  \cite{IH}
        where $\pi$, bare-$\rho$ and bare-$\sigma$ are all basic degrees
         of freedom.

\noindent
{\it Model D}: The leading order chiral lagrangian ${\cal L}_2$,
        by which the $\sigma$-pole is generated dynamically
        while the physical $\rho$  is not \cite{CGL}.

        To incorporate the effect of chiral symmetry restoration
       in these models, we replace the
        pion decay constant $f_{\pi}$ by an in-medium decay constant
       $f_{\pi}^*$ which is expected
        to decrease in nuclear matter \cite{HK85}.
        In the  $\sigma$-models,
         this replacement is approximately justified in the
         mean-field level where the
          effect of the Fermi-sea is absorbed in the redefinition of
       $f_{\pi}$ \cite{JHK}.
     In general, there arises two decay constants in the medium:
   $f_{\pi, t}^*$ which
    is related to the temporal component of the axial current,  and
   $f_{\pi, s}^*$ which
    is related to the spatial component of the axial current.
   In the qualitative study given below,
  we will not consider this complication.
   For more discussions on $f_{\pi}^*$,
   see recent papers \cite{PIA}.

       As for the  parameters such as the masses of bare-$\rho$ and
       bare-$\sigma$ and their couplings to $\pi$,
       we keep them  density
       independent,  partly because it is a simplest possible choice and
        partly because their density dependence is not known.
        Nevertheless, we shall show that, in all the above  models,
          there arise {\em simultaneous} softening of $\sigma$ and $\rho$
          driven by the decrease of $f_{\pi}^*$.
          This is the first strong indication that both the
          scalar and vector mesons serve as chiral soft modes.
         To extract essential physics of the soft modes
         without complication due to the finite
        pion mass,  we will work in the chiral limit below.


       The $\pi$-$\pi$ scattering amplitude and associate
        phase shift in the vacuum ($f_{\pi}$=93 MeV)
        in {\it Model C} has been studied in \cite{IH}
      by using the $N/D$ method.
        We briefly recapitulate its essential parts which are necessary
       for the analyses in this paper.
       The invariant amplitude for the elastic process
        $\pi^a+\pi^b\rightarrow \pi^c+\pi^d$
        is written as
        ${\cal M}_{abcd}(s,t)=A(s,t)\delta_{ab}\delta_{cd}+$
         (crossing  terms),
        where $a,b, \cdots $ denote
        isospin indices and $s,t$ are the Mandelstam variables.
        The tree-level amplitude  $A^{\rm tree}(s,t)$
       is parameterized by the contact $\pi$-$\pi$
       interaction from ${\cal L}_2$
        together with  the exchange of bare-$\sigma$ in the
     $s$-channel
       and the bare-$\rho$ exchange  in $t,u$-channels \cite{IH}.
        Then the partial
        wave amplitude $a_{_{IJ}}$  has a general form
\begin{eqnarray}
a_{_{IJ}}^{\rm tree}(s)=b_{_{IJ}}s+\sum_{\alpha=\sigma, \rho}
b'_{_{\alpha IJ}} f_\alpha(s/{\bar m}^2_{\alpha}) .
\label{subeq:tree}
\end{eqnarray}
       Here ${\bar m}_{\sigma (\rho)}$ is the mass
        of the  bare-$\sigma$(-$\rho)$.
        The first term in the right hand side  of
        eq.(\ref{subeq:tree}) is a {\em model-independent} part
        where $b_{_{IJ}}$ is the low energy
         constant solely determined  by chiral
       symmetry;
\beq
b_{00} = \frac{1}{16\pi f_{\pi}^2}, \
b_{20} = -\frac{1}{32\pi f_{\pi}^2}, \
b_{11} = \frac{1}{96\pi f_{\pi}^2}.
\eeq
       The second term in
     (\ref{subeq:tree})
   is a {\em model-dependent} part
       where the coefficient $b'_{_{\alpha IJ}}$
        is proportional to $g_{\alpha}^2$ with
       $g_{\alpha}$ being a strength of the coupling
       of $\alpha$ with two pions.
       $f_{\alpha}(s/{\bar m}^2_{\alpha})$
       behaves as $O(s^2)$ for $s \to 0$.
       We take ${\bar m}_{\sigma}$=$\infty$ ({\it Model A}),
       ${\bar m}_{\rho}$=$\infty$({\it Model B}),
           and ${\bar m}_{\sigma, \rho}$=$\infty$ in {\it Model D}.
           For {\it Model C}, we take  the same constraints as
    ref.\cite{IH},
      namely, $g_\sigma$=$g_\rho$ and ${\bar m}_{\sigma}={\bar m}_{\rho}$.

       In the $N/D$-method, the full amplitude is written 
 \beq
 a_{_{IJ}}= \frac{N_{IJ}}{D_{IJ}},
\eeq
 where $N_{{IJ}}$ ($D_{{IJ}}$) has a
       left- (right-) hand cut in the complex $s$-plane.
        The elastic unitarity  implies
 ${\rm Im}D_{{IJ}}(s>0) = - N_{{IJ}}$.
     For the numerator function, we take
  $N_{IJ}(s)$=$a_{_{IJ}}^{\rm tree}(s)$
     together with
      $D_{{IJ}}(0)=1$, which are consistent with
     the boundary condition $a_{_{IJ}}(s \rightarrow 0)
      \rightarrow b_{_{IJ}} s $ obtained from chiral symmetry.
     Then the dispersion relation for $D_{{IJ}}(s)$ reads
\begin{eqnarray}
            D_{{IJ}}(s) =  \frac{1}{\pi} \int_{0}^{\infty}ds'
            \frac{-a^{\rm tree}_{_{IJ}}(s')}{s'-s - i \epsilon} + ({\rm
subtraction}).
\end{eqnarray}
Since $a_{_{IJ}}^{\rm tree}(s \rightarrow \infty) \propto s$,
two subtractions are necessary to make $D_{{IJ}}(s)$ finite. Namely
       two unknown parameters appear. One of them can be fixed
    by  $D_{IJ}(0)=1$ mentioned above. Another one, which
      is written  as $d'_{_{IJ}}(\mu)$ below,
      should be determined empirically at some energy scale
$\mu$. Thus the final form of the denominator function  reads
\begin{eqnarray}
        D_{{IJ}}(s) &=& 1-d'_{_{IJ}} (\mu) s \cr
        &+&\frac{1}{\pi}
        \left( b_{_{IJ}}s\ln\frac{-s}{\mu^2}
                -\sum_{\alpha=\sigma,\rho} b'_{_{\alpha IJ}}d_\alpha(s/{\bar
        m}_{\alpha}^2) \right). \label{subeq:1}
\end{eqnarray}
        Here $d_{\alpha}(x)$ is obtained by the
        dispersion integral of $f_{\alpha}(x)$ in (\ref{subeq:tree}).
        Note that  $D_{IJ}(s)$ is $\mu$ independent as a whole.

        For our purpose, we need (i) to
        find complex poles of $a_{_{IJ}}(s)$ (or equivalently
        the complex zeros  of $D_{{IJ}}(s)$)
        in the 2nd Riemann sheet of the $s$-plane, and (ii) to find
        trajectories of those poles as a function of $f_{\pi}^*$.
        In eq.(\ref{subeq:1}), $\log(-z)$ and $d_\alpha(z)$ have
        multi-valued structure. $d_\alpha(z)$ contains both  $\ln (-z)$
        and  the Spence function ${\rm  Sp}(z)=-\int_0^z dy\ln(1-y)/y$
       \cite{IH}.
        The former (latter) has a branch cut along the real axis
        for ${\rm Re}z \ge 0$ (${\rm Re}z  \ge 1$).  Then the
        analytic continuation of the Spence function to the 2nd  sheet reads,
\begin{eqnarray}
       {\rm Sp}^{\rm II}(z) = {\rm Sp}^{\rm I}(z) +
       2\pi i \cdot  {\rm Ln}(z),
     \label{subeq:2}
\end{eqnarray}
        where $z$ is located in the lower half plane and
        ${\rm Ln}(z) \equiv \ln |z| +i \theta$ $(-\pi < \theta \le \pi)$.
        The superscript I (II) indicates the 1st (2nd) sheet \cite{spence}.


Here we mention briefly the determination of the
      unknown parameters ($d_{_{IJ}}'$,
       $g_{\alpha}$ and ${\bar m}_{\alpha}$) in each model.
      In {\it Models A, B} and {\it C},
      $g_\rho$ is fixed to be
      $g_{\rho}^2/4\pi = 2.72$ for reproducing the $\rho$-width.
       In {\it Model B} and {\it C}, $g_\sigma$ is assumed to be
     the same as $g_\rho$ following Ref. \cite{IH}.
       Remaining parameters $d'_{11}$ and ${\bar m}_\alpha$ are adjusted
      by making a global fit to
      the experimental phase shift in  the $I$=$J$=1 channel up to
      1 GeV \cite{rho}. Then,
      $d'_{00}$ and $d'_{20}$  are determined uniquely from $d'_{11}$
      by the constraints obtained from the matching to the
       $O(p^4)$  chiral lagrangian for small $s$ \cite{IH,MAT}.
      For {\it Model D},  $d_{_{IJ}}$ is directly extracted
      from  the  coefficients $L_{1,2}^r(\mu)$ in the $O(p^4)$  chiral
      lagrangian.   The $I$=$J$=1 phase shift is not reproduced
      in {\it Model D}, since only ${\cal L}_2$ is considered \cite{CGL}.
      The results are listed in TABLE~\ref{tab:parameter}
      for each model. We take the standard choice $\mu$=0.77 GeV
       in determining the parameters.
     We have checked that the experimental
      phase shift in the ($I$,$J$)=(0,0) and
      ($I$,$J$)=(2,0) channels is reproduced quite well within the
      experimental error-bars  in {\it Model B} and {\it C} below 0.9 GeV.
       {\it Model A} and {\it D} also show a qualitatively reasonable fit in
        the above channels at low energies. However,
       {\it Model A} underestimates
         attraction  for ($I$,$J$)=(0,0) above 0.6 GeV.
       Also, {\it Model D} overestimates the attraction and repulsion
       above 0.4 GeV for ($I$,$J$)=(0,0)  and  ($I$,$J$)=(2,0),
    respectively.

\begin{table}
\caption{\label{tab:parameter}
The bare mass ${\bar m}_{\alpha}$
        and $d_{11}'$ extracted from the global fit of
      the phase shift in the $I$=$J$=1 channel up to
      1 GeV \cite{rho}.
      For {\it Model D}, we use $L_1^r(\mu)=0.4 \times 10^{-3}$ and
      $L_2^r(\mu)=1.4 \times 10^{-3}$ \cite{Para}
        to extract $d'_{_{IJ}}$ at $\mu = 0.77$ GeV.
          $d_{00}'$ obtained from
       $d_{11}'$ is also shown.}
\begin{ruledtabular}
\begin{tabular}{rrr|r}
Model  &${\bar m}_{\alpha} $(GeV)& $d'_{11}$(GeV$^{-2}$) &
$d'_{00}$(GeV$^{-2}$) \\
\hline
A & ${\bar m}_{\rho}$=0.774           & $-$0.385   &     0.171 \\
B & ${\bar m}_{\sigma}$=0.838         & 1.66       &      $-$3.04\\
C & ${\bar m}_{\rho, \sigma}$=0.778   & $-$0.342   &      $-$0.0855\\
D & $-$                               & 0.277      &      2.19\\
\end{tabular}
\end{ruledtabular}
\end{table}


       Before making numerical analysis of
       $D_{IJ}(s)$=0, let us first discuss its
       solution for  small values of $s$ where
       eq.(\ref{subeq:1}) may be
        approximated as
 \beq
 D_{IJ} (s) \simeq 1+(b_{_{IJ}}/\pi) s \ln (-s/M^2),
\eeq
        in all four models. Here we have neglected $O(s^{n \ge 2})$ terms
      and $M$ is a $\mu$-independent scale defined as
    $M^2 \equiv \mu^2 e^{d'(\mu)\pi/b}$.
     Then a solution on the 2nd Riemann sheet in ($I$,$J$)=(0,0), (1,1)
    channels reads
\begin{eqnarray}
      s_{_{IJ}} & = & -(\pi/b_{_{IJ}} ) \cdot
     \left[
      W_{-1} ( \pi /b_{_{IJ}} M^2 )
     \right] ^{-1}
    \label{lambert-1}  \\
              & \rightarrow &  \frac{- F_{_{IJ}}^{*2} }
                        { {\rm Ln}\left[ - (F_{_{IJ}}^{*2}/M^2)
                         /{\rm  Ln}(F_{_{IJ}}^{*2}/M^2) \right]
                           - i \pi } ,
      \label{lambert}
\end{eqnarray}
     where $W_{-1}(z)$ is  the $(-1)$-th branch of the
      Lambert $W$ function \cite{LWF1}.  From eq.(\ref{lambert-1})
       to eq.(\ref{lambert}),
     we have used an asymptotic expansion
      of $W_{-1}(z)$ valid for small and positive $z$   \cite{LWF2}.
    Also,   $F_{_{00}}^* = 4 \pi f_{\pi}^*$ and
    $F_{_{11}}^* = \sqrt{6} F_{_{00}}^*$.
       Eq.(\ref{lambert}) explicitly shows  that
        resonance poles appear both
       in the $\rho$ and the $\sigma$ channels, and   are softened
       {\it in tandem} as $f_{\pi}^*$ decreases in nuclear medium.

\begin{figure}
\includegraphics[scale=0.5]{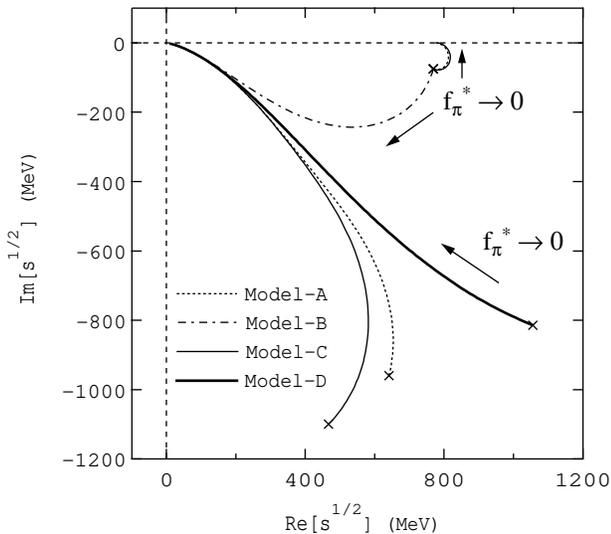}
\caption{\label{fig:11pole} The shift of the pole position in the
$I$=$J$=1 channel associated with the decrease of $f_{\pi}^*$.
Two kinds of flows exist: one toward the origin and the other
       toward ${\bar m}_{\rho}$ on the real  axis.
       Crosses are the pole positions in the vacuum.}
\end{figure}

\begin{figure}
\includegraphics[scale=0.5]{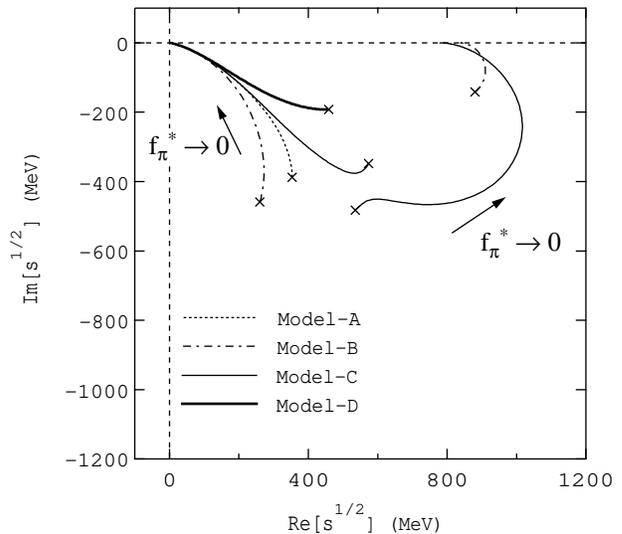}
\caption{\label{fig:00pole} The shift of the pole position in the
$I$=$J$=0 channel associated with the decrease of $f_{\pi}^*$.
Two kinds of flows exist: one toward the origin and the other
       toward ${\bar m}_{\sigma}$ on the real  axis.
Crosses  are the pole positions in the vacuum.}
\end{figure}

     For $f_{\pi}^* \sim f_{\pi}$, the small $s$
       approximation cannot be justified and more complicated
      behavior of the poles arises.
       To keep track of the trajectory of the poles
       for wide range of $f_{\pi}^*$,
      we show  numerical solutions of  eq.(\ref{subeq:1})
       for the
        $I$=$J$=1 channel  (Fig.~\ref{fig:11pole})  and
        for the $I$=$J$=0 channel (Fig.~\ref{fig:00pole}).
        In these figures, the crosses indicate
       the position of the poles in the vacuum ($f_{\pi}^*$=$f_{\pi}$).
        A common feature in {\it Models A, B and C} in the vacuum
       is that there always
       exists a narrow $\rho$  ($\sqrt{s_\rho} \simeq 769 - 76 i$ MeV)
       and a broad/low-mass $\sigma$, no matter whether
       bare resonance are introduced or not.
       This is a kind of bootstrap situation discussed in \cite{BAS}.
       Note also that, when  the bare resonance is introduced,
       two complex poles appear as a result of an interplay between the
       bare resonance and the dynamically generated one.
       This can be seen in Fig.~\ref{fig:11pole} for {\it Models A and C},
      and
        in Fig.~\ref{fig:00pole} for {\it Models B and C}.
       In {\it Model D} where bare resonances are not introduced,
       a broad and low-mass $\sigma$ is dynamically generated
       while   narrow $\rho$ does not appear.

       As $f_\pi^*$ decreases from its vacuum value,
        we find two types of trajectories
       in Figs.~\ref{fig:11pole}-\ref{fig:00pole};
       those moving toward the origin and
       those moving toward  ${\bar m}_{\rho(\sigma)}$ on the real axis.
       The former trajectories in   $I$=$J$=1 and $I$=$J$=0  channels
       are model independent and correspond to the
       universal softening evaluated in Eq.(\ref{lambert}).
        Since the $\pi$-$\pi$ scattering amplitude is mainly
       affected by the  poles close to the real axis,
       such soft mode gives a dominant contribution
       to the low-energy amplitude for $f_{\pi}^*  /f_{\pi} \ll 1$.

       On the other hand, when the system is close to the
       vacuum state  ($f_{\pi}^* / f_{\pi} \sim 1$),
       the above  universality between $\rho$ and $\sigma$ breaks down.
       In fact,  in the $I$=$J$=1 channel,
       the narrow $\rho$ resonance plays a dominant role
        and  would-be soft mode is far away from the real-axis.
       In the $I$=$J$=0 channel, on the contrary,
       the broad and low-mass  $\sigma$ keeps playing a
       role of the soft mode all the time.

\begin{figure} [t]
\includegraphics[scale=0.5]{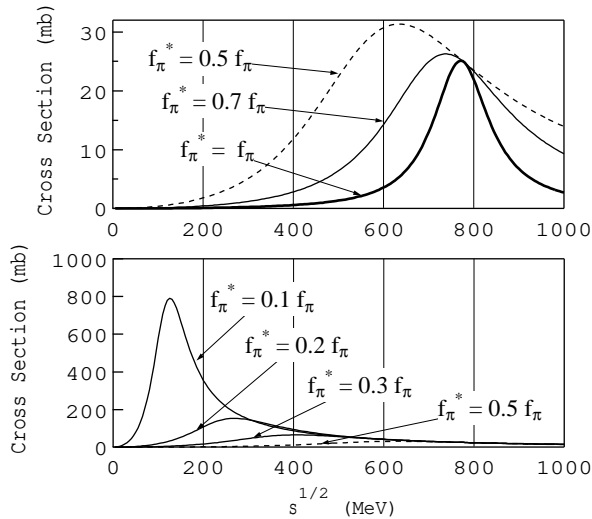}
\caption{\label{fig:test11}
       The in-medium $\pi$-$\pi$ cross section in the $I$=$J$=1 channel
       for {\it Model-B}.
       The upper (lower) panel shows the case
       for  $0.5 f_\pi \le f_\pi^* \le f_\pi$
        ($0.1 f_\pi \le f_\pi^* \le 0.5 f_\pi$).
}
\end{figure}

\begin{figure} [b]
\includegraphics[scale=0.5]{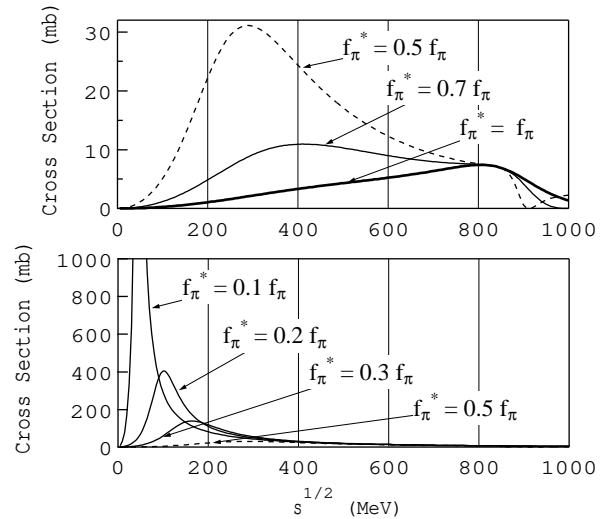}
\caption{\label{fig:test00}
       Same with Fig. \ref{fig:test11} for
       the $I$=$J$=0 channel for {\it Model-B}.
       \mbox{ }
}
\end{figure}

       The similarity and difference between the two channels
       for different values of $f_{\pi}^*$
       can be  seen also  by comparing the
       in-medium $\pi$-$\pi$ cross section calculated from $a_{_{IJ}}(s)$.

        The upper panel of Fig.\ref{fig:test11} (Fig.\ref{fig:test00})
        represents the in-medium
        cross section in {\it Model B} for
        $0.5 f_\pi \le f_\pi^* \le f_\pi$
        in the $I$=$J$=1 ($I$=$J$=0) channel.
        We find that there is  a moderate softening + broadening
        of the $\rho$-resonance at $f_{\pi}^* = 0.7 f_{\pi}$, which is
        itself an interesting behavior in relation to the dilepton data
        in \cite{CERES,E325}. However, even more
        significant change can be seen for the $\sigma$-resonance
        at the same value of $f_{\pi}^*$.

       The lower panel of Fig.\ref{fig:test11} (Fig.\ref{fig:test00})
       represents the in-medium
       cross section in {\it Model B} for
       $0.1 f_\pi \le f_\pi^* \le 0.5 f_\pi$
       in the $I$=$J$=1 ($I$=$J$=0) channel.
       We find that sharp peaks develop in
       both channels as $f_{\pi}^* \rightarrow 0$. They are the
       direct consequence of the soft modes located near the origin
        for small $f_{\pi}^*$ in Figs.~\ref{fig:11pole}-\ref{fig:00pole}.
       Namely, both $I$=$J$=1 and $I$=$J$=0
       channels are good probes of chiral restoration in the medium if
       the chiral  restoration  is substantial.
       Although we have taken {\it Model B} as an example in Fig.3-4, we
       have
       checked  that qualitative conclusions are the same for other
       models.


In summary, we have studied
        complex poles of the  in-medium $\pi$-$\pi$ scattering amplitude
       in the $I$=$J$=0 and $I$=$J$=1 channels to explore
      possible relations  between $\sigma$ and $\rho$
      in nuclear matter.
      The $N/D$ method is applied to four types of chiral models
      and  the chiral restoration is modeled by the decrease of $f_\pi^*$.
      We have found universal complex poles  in  both channels
      moving  in tandem toward the origin if $f_{\pi}^*$ is
      sufficiently small. On the other hand, if $f_{\pi}^*$ is
      not far from its vacuum value, interesting non-universal behavior
    arises and the two resonances act  rather differently.

      Inclusion of the  finite pion mass, the medium effect beyond the
      mean-field approximation (such as the
     coupling to the particle-hole excitations of the
    Fermi-sea and the difference of $f_{\pi,t}^*$ and
   $f_{\pi,s}^*$ mentioned before) are future problems to be examined.
  Possible connection to other theoretical  approaches \cite{HKR}
 should be also studied.

This work is  partially supported by the Grants-in-Aid of
the Japanese Ministry of Education, Science and Culture
(No. 12640263 and 12640296).
A.H. is supported by JSPS Research Fellowship for Young Scientists.

\end{document}